# Development of Parallel Plate Avalanche Counter PPAC for BigRIPS fragment separator

H. Kumagai, T. Ohnishi, N. Fukuda, H. Takeda, D. Kameda, N. Inabe, K. Yoshida, T. Kubo

*RIKEN Nishina Center, RIKEN, 2-1 Hirosawa, Wako, Saitama 351-0198, Japan*

**Abstract**

We have developed a position-sensitive Parallel Plate Avalanche Counter (PPAC), which has been used as a focal plane detector in the BigRIPS fragment separator and the subsequent RI-beam delivery lines at the RIKEN Nishina Center RI Beam Factory. The PPAC detector plays an important role not only in the tuning of the separator and delivery lines but also in the particle identification of rare isotope (RI) beams. The PPAC detector has a sensitive area of 240 mm × 150 mm, and the position information is obtained by a delay-line readout method. Being called double PPAC, it is composed of two full PPACs, each measuring the particle locus in two dimensions. High detection efficiency has been made possible by the twofold measurement using the double PPAC detector. The sensitivity uniformity is also found to be excellent. The root-mean-square position resolution is measured to be 0.25 mm using an α source, while the position linearity is as good as ±0.1 mm for the detector size of 240 mm. Characteristics, operating principles, specifications, performance and issues of the PPAC detector are presented, including its signal transmission system using optical fiber cables.

## 1. Introduction

Since November 1989, position-sensitive Parallel Plate Avalanche Counters (PPACs) have been important focal plane detectors for detecting heavy ions at the RIPS fragment separator [1] at RIKEN. Improved PPAC detectors have been developed and used at the

BigRIPS fragment separator [2-4], which was commissioned in March 2007 at the RIKEN RI Beam Factory (RIBF) [5]. The PPAC detectors play a very important role in not only the tuning of the fragment separator and subsequent RI-beam delivery lines but also the production and delivery of rare isotope (RI) beams in which particle identification is made in an event-by-event mode. Adopting a delay-line readout method, they measure positions and angles (or trajectories) of RI beams at the foci, by which the beam diagnostics and ion-optical tuning are made possible. Furthermore the PPAC detectors are used for Bρ (magnetic rigidity) determination in the particle identification of RI beams, in which trajectory reconstruction is employed to achieve high Bρ resolution and hence excellent particle identification power [6,7]. The PPAC detector developed for the BigRIPS separator has a larger sensitive area, and consists of two layers, each measuring the particle locus in two dimensions. Acknowledging its structure, it is called "double PPAC", and offers higher detection efficiency compared to a "single PPAC". Transmitting data over fiber-optic cables, the double PPAC detector renders excellent performance as a focal plane detector.

One of the most important features of the PPAC detector is that the amount of substance is quite small in comparison with other kinds of position-sensitive gaseous detectors, such as a multi-wire proportional counter [8] and a multi-wire drift chamber [8]. The amount is approximately 30 mg/cm$^2$, which is less than 1/10 of the amount of other detectors, thereby perturbing the production and delivery of RI beams significantly less, including measuring event-by-event trajectories. Furthermore, the perturbation of the beam is even less because no wires are used. Another important feature is that the PPAC detectors are durable and their maintenance is relatively easy because of their simple structure.

In the present paper, we present the characteristics, operating principles, specifications, performance and issues of the PPAC detector that we developed. Our independently developed analog light transmitter and receiver, which are used for the signal transmission with optical fiber cables, are also introduced.

## 2. History of PPAC detectors and operating principles

The PPAC detector was first used in 1952 by Christiansen [9]. Recognition of their superior performance as heavy ion detectors has made the PPAC detectors popular since around 1975 [10-12]. A prototype one-dimensional PPAC was built at RIKEN for the first time in 1984, and its first use in an experiment was the measurement of the charge state distribution of heavy ions in 1985 [13], an atomic physics experiment.

Two-dimensional position sensitive PPAC detectors were built and have been used as focal plane detectors in the RIPS separator since 1989.

The PPAC detector consists of two parallel thin electrode films separated by 3–4 mm and is filled with 3–50 Torr of gases such as isobutene ($C_4H_{10}$) and perfluoropropane ($C_3F_8$), as shown in Fig. 1. When a voltage corresponding to a few hundreds of volts per millimeter is applied between the anodes and cathodes, ionized electrons from incident heavy ions immediately cause an electron avalanche. Because there is no time delay before the avalanche occurs and the electrons move at high mobile velocity (mobility), the resulting signals have good timing properties, with rise and fall times of a few nanoseconds, as compared with other types. For instance, the cylindrical proportional counter shown in Fig. 2 shows the time delay before an avalanche starts, because the ionized electrons have to move near the sensitive wire.

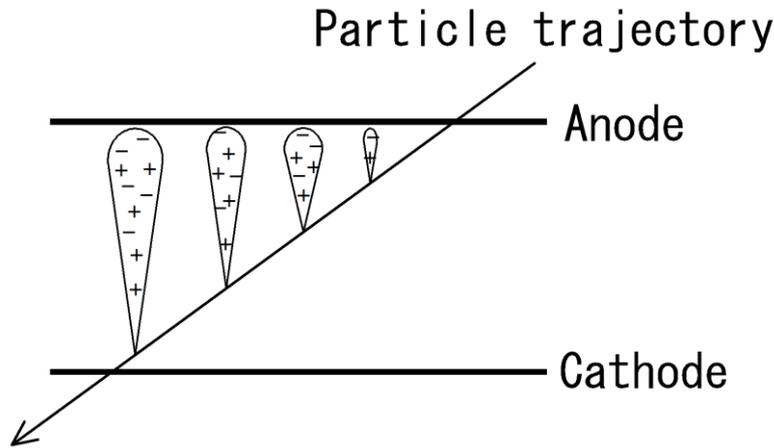

Fig. 1. Operating principles of PPAC.

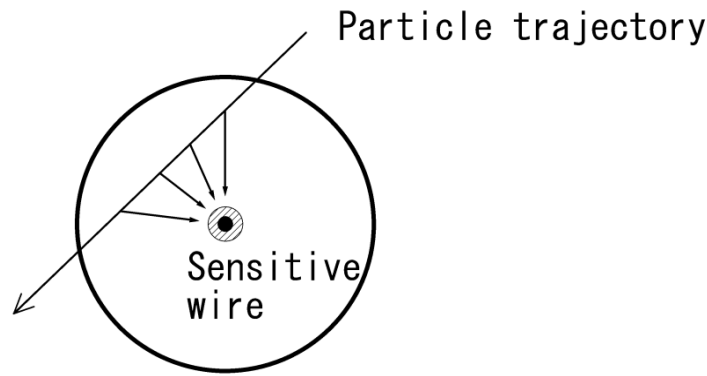

Fig. 2. Operating principles of cylindrical proportional counter.

The avalanche here is a Townsend avalanche [14-16], in which the increasing rate of electrons is proportional to the number of electrons, as described by the equation

$$\frac{dN}{dz} = \alpha N,$$

where N, z, and α represent the number of electrons, the distance along the electrode gap, and the proportional coefficients. Solving this differential equation and substituting the initial conditions $z = 0$ and $N = N_0$, where $N_0$ is the number of primary electrons (initially ionized electrons), give

$$N = N_0 \exp(\alpha z), \qquad (1)$$

which is the well-known Townsend avalanche equation [14-16]. Here, *α* represents the first Townsend coefficient, which is determined by the type of gas, pressure, and electric field.

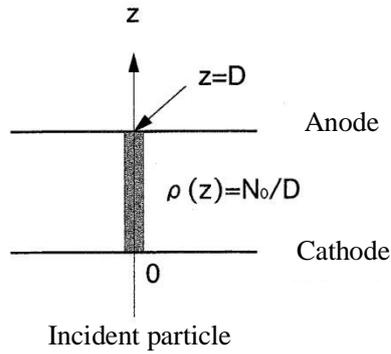

Fig. 3. Distribution of primary electrons in a PPAC detector. ρ(z) represents the density of the number of primary electrons along the particle trajectory.

Equation (1) holds when $N_0$ primary electrons start at a single point. In a PPAC detector, however, the primary electrons are uniformly distributed along the trajectory of a particle, as shown in Fig. 3. Therefore, the integrated number of avalanche electrons N becomes

$$N = \int_0^D \frac{N_0}{D} \exp(\alpha z) dz$$

$$= \frac{N_0}{\alpha D} \{\exp(\alpha D) - 1\},$$

where D represents the gap width between the anode and cathode electrodes. Since

$$\exp(\alpha D) \gg 1,$$

$$N = \frac{N_0}{\alpha D}\exp(\alpha D). \qquad (2)$$

Comparing Eq. (2) with Eq. (1) when $z = D$ yields

$$N_0 \rightarrow \frac{N_0}{\alpha D},$$

indicating that the number of primary electrons is multiplied by $1/\alpha D$. In other words, the number of effective primary electrons is $1/\alpha D$ times the number of electrons $N_0$ that are formed between the electrodes.

The coefficients $D = 4$ mm and $\alpha = 3.0$ mm$^{-1}$, which are typical parameters for our PPAC detectors, result in $1/\alpha D \approx 0.08$. Therefore, effectively only 8% of the electrons are used. This is one reason that PPAC detectors are less sensitive than cylindrical proportional counters.

## 3. Position measurement using PPAC detectors

Position measurement in a PPAC detector is conducted by the charge division method or by the delay-line readout method, in which a multi-strip cathode electrode is used to measure the induced charge distribution. One can read out the signals for each strip, as in a multi-wire proportional counter. However, this is not so effective because the output signal in a PPAC detector is small.

### 3.1. Charge division method

To employ this method, the electrode strips are connected to resistor arrays, and the position is measured from the ratio of the charges appearing at either end. Because this structure is relatively easy to make, the method was initially adopted at the RIPS separator. This method allows us to fully utilize all the charge output because the signals originating from cations (ionized positive ions) are used. Furthermore, filter amplifiers with good noise characteristics can be used as the main amplifier, so that highly sensitive measurement can be made possible. This is the advantage of this method. In fact, we could measure the position distribution of a 100-MeV proton beam at the momentum dispersive focus F1 [1] in the RIPS separator.

However, we no longer use the charge division method at RIKEN because of the following disadvantages. The preamplifier is charge sensitive and the time constant of the main amplifier is long at 0.5 - 2 μs, which prohibits measurement at high counting rates because signal pile-up occurs at a few thousand Hz. Furthermore, the dynamic range is narrow, because one can use only the region with good linearity. Hence, the range of atomic number Z that can be measured is also narrow.

*3.2. Delay-line readout method*

Here, the electrode strips are connected to multi-tapped delay lines, and the position is determined from the time difference between signals appearing from either end. The signal width is as short as several nanoseconds because fast signals generated from avalanche electrons are used, allowing the PPAC detector to handle fast measurements at a few million Hz. In addition, the dynamic range is wide because somewhat poor amplitude properties are tolerated as long as the timing properties are acceptable. Furthermore, the effects of δ rays and multiple hits can be removed using characteristics unique to the delay-line readout.

All the PPAC detectors in the RIPS and BigRIPS separators use the delay-line readout method because of these benefits. However, the drawback is its low sensitivity because fast signals are generated by only 9% of the total charge. Although all the electrons generated by an avalanche are absorbed at the anode electrode, they cannot immediately leave the electrode because slow cations still stay near the anode surface. The centroid of the cation charge distribution is positioned at $1/\alpha \approx 0.3$ mm from the anode electrode, corresponding to 3.7 mm from the cathode electrode, if D = 4 mm. In this case, 91% of the electrical flux lines point to the anode electrode. The rest (9%) points to the cathode electrode. The electron charge corresponding to the latter can be extracted as a fast signal. Hence only the initial rise in the PPAC output signal can be used for the fast signal.

Figure 4 shows an output waveform from a charge-sensitive preamplifier connected to the anode electrode, which was measured using an oscilloscope. The sharp rise in the waveform indicates that part of the avalanche electrons quickly left the anode, and the subsequent slow rise comes from the cations moving toward the cathode electrode. It can be seen in the figure that the amplitude of the initial rise is just 8.7% of the total output.

**4. Fabrication of the delay line**

The high performance of the delay-line PPAC detector at RIKEN comes primarily from that of our independently developed (proprietary) delay line. Here we present the design principles and performance of our delay line.

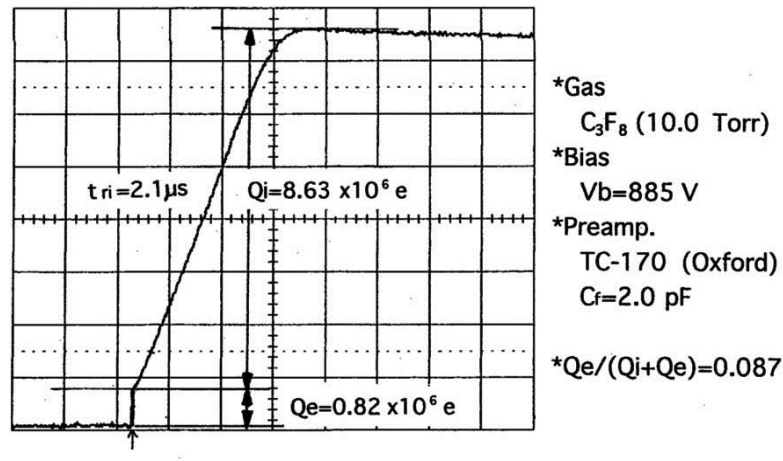

Fig. 4. Output waveform from a charge-sensitive preamplifier connected to the anode electrode of a PPAC detector, which was measured using an oscilloscope. The horizontal and vertical grids correspond to 1 μs and 200 mV, respectively.

Although coaxial cables or electrode strips can be used as a delay line, we chose to fabricate lumped delay lines by ourselves because they can be more compact and the delay time can be adjusted. We considered but decided against using commercially available lumped delay lines, because the number of intermediate taps is too small such as 20 or less, and the dispersion of the delay time among the taps is as large as ±20 %. Moreover the uniformity of the delay time gets worse at the connection point of the delay lines.

*4.1. Guidelines for design*

The guidelines in designing our delay-line PPAC detector are summarized as follows:
- Delay time step: approximately 2 ns, corresponding to half the rise time of the PPAC signals.
- Pitch of cathode strips: approximately 2 mm, corresponding to the half width at half maximum of the charge distribution that is induced on the cathode surface.
- Interval between intermediate taps of the delay line: approximately 2 mm, the same as the pitch of cathode strips.
- Characteristic impedance: $Z_0 = 50\ \Omega$.

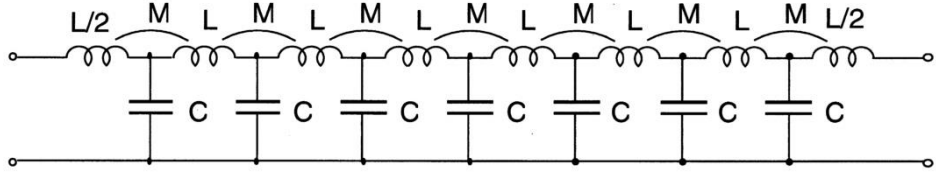

Fig. 5. Equivalent circuit of the lumped delay-line in the PPAC detector: series m-derived filter delay line. C, L, and M denote capacitance, self-inductance, and mutual-inductance, respectively.

*4.2. Parameters of LC lumped delay line*

An equivalent circuit of the lumped delay-line used in our PPAC detector is shown in Fig. 5. The parameters of the delay line, such as the delay time $t_d$, characteristic impedance $Z_0$, and coupling constant $K$ that optimizes the linearity, are analytically given as follows [17].

$$t_d = 1.20\,\text{n}\sqrt{LC}$$

$$Z_0 = \sqrt{\frac{L}{C}}$$

$$K \equiv \frac{M}{L/2} = 0.234$$

where n represents the total number of delay-line steps. The C, L, and M parameters in the lumped delay line are determined according to these conditions. However, some components, for instance, inductance coil cradles, polyurethane enameled wires, and chip capacitors, do not offer perfectly matched parameters, because they are obtained off the shelf.

The inductance coils that we could make using available components and that most closely match parameters are those using 0.3-mm-diameter polyurethane enameled wires wound around a 2-mm-diameter Bakelite rod. Lead wires are soldered every eight turns and 39-pF capacitors are used. The resulting delay line parameters are given as follows:

- $L = 0.09$ μH (calculated)
- $K = 0.287$ (calculated)

- $C = 39$ pF
- $Z_0 = 53$ Ω (measured) (calculated: 48 Ω)
- Delay-line pitch (strip pitch) = 2.55 mm (strip width = 2.4 mm and strip gap = 0.15 mm)
- Number of strips = 40 strip
- Length of delay line = 100 mm
- Total delay time $t_d$ = 2.04 ns × 40 steps = 81.6 ns (measured, average delay time of one pitch = 2.04 ns)

The last three parameters are those in the case of a 100-mm PPAC detector. The inductance coils and chip capacitors were mounted on a 2-mm-thick double-sided G10 printed circuit board with 2.55-mm-pitch strip electrodes to form the delay line. The strip electrodes of the delay-line board electrically connect the delay-line taps to the cathode strips. The width and pitch of the delay-line strips exactly accord with those of the cathode strips of the PPAC detector.

## 5. Configuration of the delay-line PPAC detectors

The electrode setup of the delay line PPAC [18] consists of $X$-axis and $Y$-axis cathode electrodes placed in parallel on both sides of an anode electrode, as shown in Fig. 6. The electrodes are formed by vacuum metal-deposition onto a thin polyester film which is pasted to the G10 frame board of 2.4-mm thickness. We stretch well the film when pasting and then heat it up using a hot-air blower to remove wrinkles. The thickness of the films is 1.5 μm for PPAC detectors with area size of 100 mm × 100 mm or smaller. In the case of those with larger size, the thickness is 2.5 μm for the anode film and 4 μm for the cathode film.

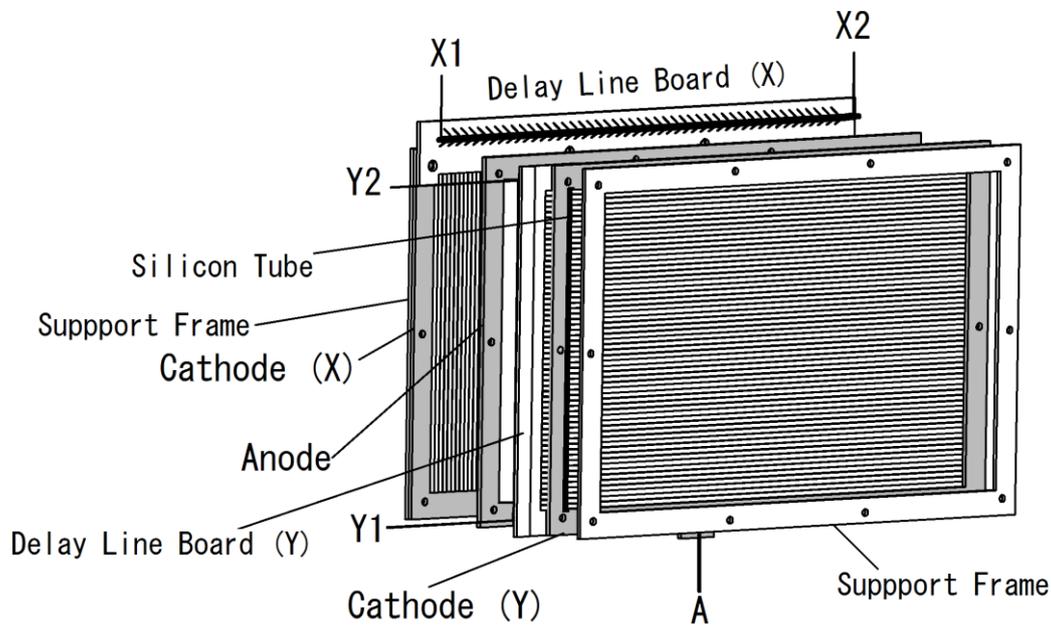

Fig. 6. Schematic view of the electrode setup in a 240mm × 150 mm PPAC detector.

We uniformly vacuum-deposited the metal such as aluminum on both sides of the anode film. In the case of the cathode electrode, the metal was vacuum-deposited in strips with a 2.55-mm pitch on one side of the film. The width of the strip is 2.4 mm and the spacing between strips is 0.15 mm. We used a multi-strip mask in front of the film to form the strips. The mask is precisely made of a stainless-steel sheet by photo-etching. The mask was tensed using small springs in order to prevent the deformation caused by the heat during the vacuum deposition. The deposition thickness is 300 Å in case of aluminum.

The delay line is attached such that the printed strips on the surface of the delay-line G10 board is in contact with the cathode strips, and the contact is secured by pushing the cathode film using an elastic silicon tube, which is placed between the cathode film and a supporting frame. In this way, the delay line does not need to be replaced when the cathode electrode film is damaged. We just change the film and the delay line can be used many times.

The gap between the electrodes is 4.3 mm, and the margin of error is set to less than ±5 μm because it significantly affects the amplitude of the output signal. The gap is kept by the spacers placed between the electrode frames. The electrodes are then positioned in a vacuum-sealed duralumin case. Window plates consisting of aluminum-deposited polyester films are placed on both sides. The standard thickness of the window film is 4 and 12 μm for small and large PPAC detectors, respectively.

We have developed and fabricated the PPAC detector called double PPAC, in which two sets of the PPAC detector shown in Fig. 6 are placed in one case, allowing twofold position measurement. This double PPAC detector allows high detection efficiency as well as providing a backup function, as discussed below. We call the PPAC detector single PPAC, when only one set of the PPAC detector is placed in one case. We used such a single PPAC detector before the BigRIPS separator was commissioned.

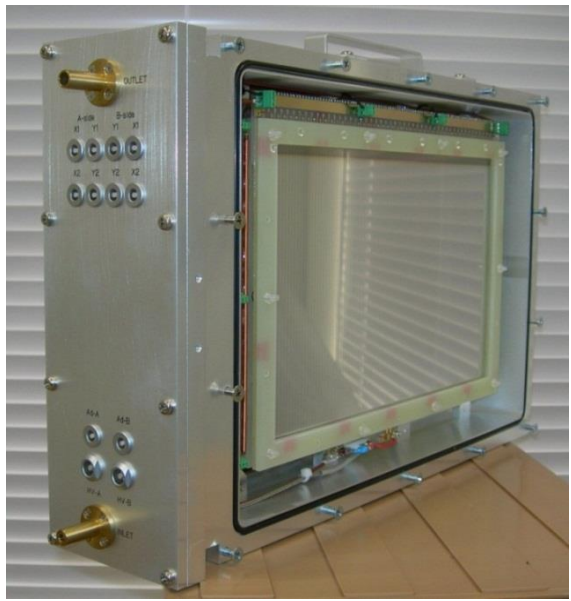

Fig. 7. Photograph of the 240 mm (X) × 150 mm (Y) double PPAC detector.

In total the following six types of delay-line PPAC detectors have been developed and fabricated so far:

- 100 mm × 100 mm, 150 mm (X) × 100 mm (Y), and 150 mm × 150 mm for the RIPS separator
- 150 mm × 150 mm, 240 mm (X) × 100 mm (Y), and 240 mm (X) × 150 mm (Y) for the BigRIPS separator

The photograph in Fig. 7 shows an external view of the 240 mm × 150 mm double PPAC detector, where the window plate is removed so that the inside is visible. The electrode seen is the $X$-axis cathode of the second PPAC, called B-side PPAC, and its delay line is located at the top of the electrode.

## 6. Evaluation of delay-line PPAC performance using an α source

### 6.1. Equivalent circuit and formula for obtaining the position

Figure 8 shows the equivalent circuit of the delay-line PPAC detector and a schematic of its operation. Fast signals induced in the cathode by an avalanche enter a delay line and split to travel in the right and left directions. Then they are output to the X1 and X2 terminals, respectively. The delay time is measured using a time-to-digital converter (TDC), where the measurement starts with the anode signal and stops with cathode signals from the X1 and X2 terminals. Denoting these two delay times as $T_{x1}$ (ns) and $T_{x2}$ (ns), the position $X$ is obtained by

$$X = K_x \times (T_{x1}-T_{x2})/2 + X_{off}, \qquad (3)$$

where $K_x$ (mm/ns) is the position coefficient, and $X_{off}$ (mm) is the offset correction. The position can also be calculated by using either $T_{x1}$ or $T_{x2}$. However, by Eq. (3), we see that delays are cancelled from the cables and circuits and the $X^2$ nonlinear component in the TDC output is removed.

The sum of the delay times $T_{sum}$, defined as

$$T_{sum} \equiv T_{x1}+T_{x2}, \qquad (4)$$

has a constant value independent of the position of the incident particle and works as an important parameter for removing the effects of $\delta$ rays and multiple hit events.

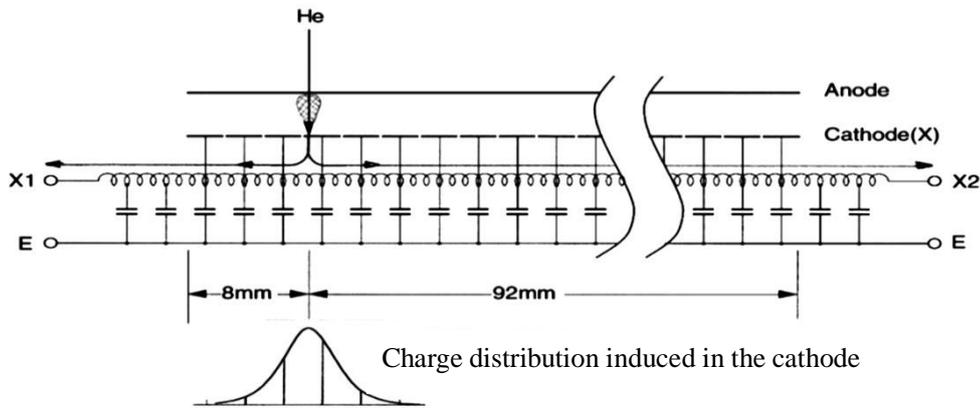

Fig. 8. Equivalent circuit of the delay-line PPAC detector.

*6.2. X-axis position spectrum from the 240 mm × 150 mm PPAC detector*

Figure 9 shows the X-axis position spectrum of α rays which are uniformly irradiated on the 240 mm × 150 mm PPAC detector, demonstrating the excellent uniformity of the detection sensitivity. An $^{241}$Am α-source was used in this measurement. Because the pitch of the cathode strips is 2.55 mm, the patterns of the strips should have been seen in the spectrum. However, they are not visible because our PPAC detectors have unique setting for the parameters and constants, as discussed below.

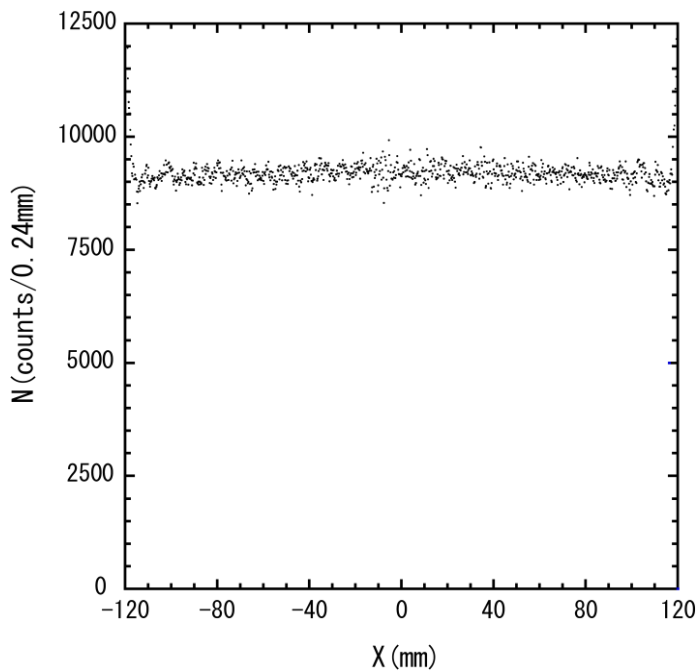

Fig. 9. X-axis position spectrum measured with α rays from an $^{241}$Am source, which are uniformly irradiated on the 240 mm x 150 mm PPAC detector.

Figure 10 shows the X-axis position spectrum measured by using a position calibration mask, which has a slit interval and width of 5 mm and 0.5 mm, respectively. The root-mean-square (r.m.s.) position resolution has been measured to be $\sigma = 0.25$ mm, demonstrating that the PPAC detector reveals enough performance to be used as a focal plane detector in the BigRIPS separator.

*6.3. Position of incident particle and waveform of delay-line output*

We investigated the reason why the pattern of the cathode strips does not appear in the uniformly irradiated position spectrum, although the pitch of the cathode strips is as large as 2.55 mm.

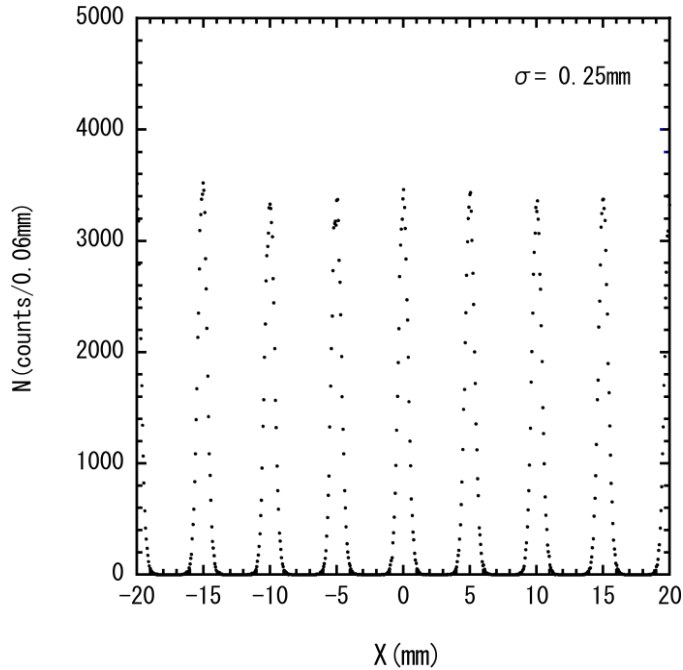

Fig. 10. X-axis position spectrum with a position calibration mask which are measured with α rays from an $^{241}$Am source.

For this purpose, we fabricated a test PPAC detector with a smaller pitch of cathode strips. The strip pitch was chosen to be 1.5 mm and the charge distribution induced on the surface of cathode electrode has been measured. A charge-sensitive analog-to-digital converter was used to measure the amount of the induced charge for each strip. Figure 11 shows the charge distribution thus obtained. The distribution can be fitted well with a Gaussian with σ = 1.80 mm. Based on this function, we calculated the amount of the induced charge for each strip when the strip pitch is 2.55 mm.

Figure 12 shows the waveform of the cathode signal that was measured using an oscilloscope for the test PPAC detector. The measured waveform could be fitted well to a Gaussian with $\sigma$ = 2.13 ns. Based on this measured waveform of the cathode signal and the measured distribution of the induced charge, we calculated the waveform for each strip when the strip pitch is 2.55 mm, and derived the total waveform of the delay-line output by summing these waveforms. The strip pitch of 2.55 mm gives rise to a time difference of 2.04 ns. We took it into consideration in deriving the total waveform.

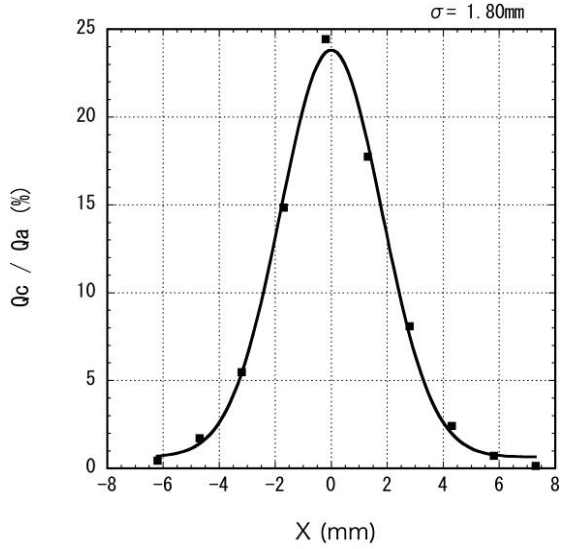

Fig. 11. Charge distribution induced in the cathode surface that was measured using the test PPAC detector with a cathode strip pitch of 1.5 mm. The solid line shows a Gaussian fit with $\sigma = 1.80$ mm.

We analyzed the difference in the waveform of the delay-line output between the following two cases: (1) a particle is incident at the center of a cathode strip and (2) incident at a gap between cathode strips. Figure 13 shows the calculated total waveform of the delay-line output in the former case along with the waveforms from five cathode strips. The total waveform is a Gaussian with $\sigma = 2.57$ ns. The calculated total waveform and the cathode strip waveforms in the latter case are shown in Fig. 14. In this case the total waveform becomes a Gaussian with $\sigma = 2.58$ ns, which is very similar to the former case, indicating that the delay-line output waveform in our PPAC detector changes very little with the incident position of a particle. This is the reason why the position spectrum shown in Fig. 9 is uniform.

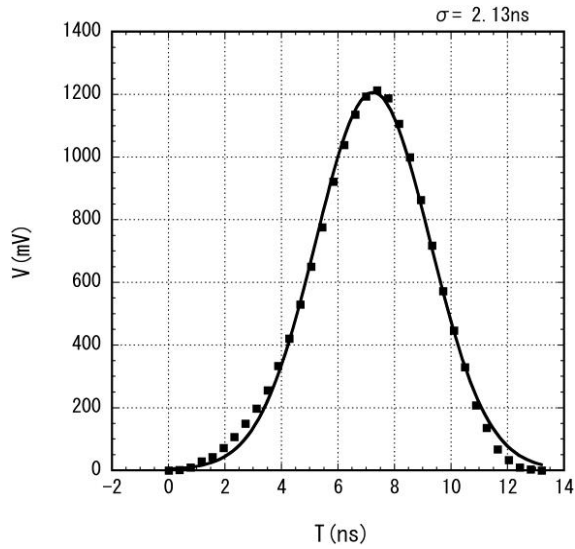

Fig. 12. Waveform of the cathode signal measured using an oscilloscope for the test PPAC detector. The solid line shows a Gaussian fit with σ = 2.13 ns.

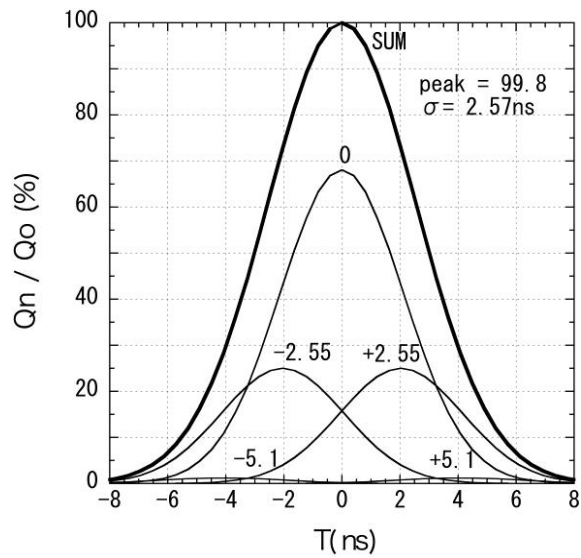

Fig. 13. Calculated output waveforms for a particle incident on the center of a strip. The Gaussian labeled by SUM is the total waveform, while others are those from strips.

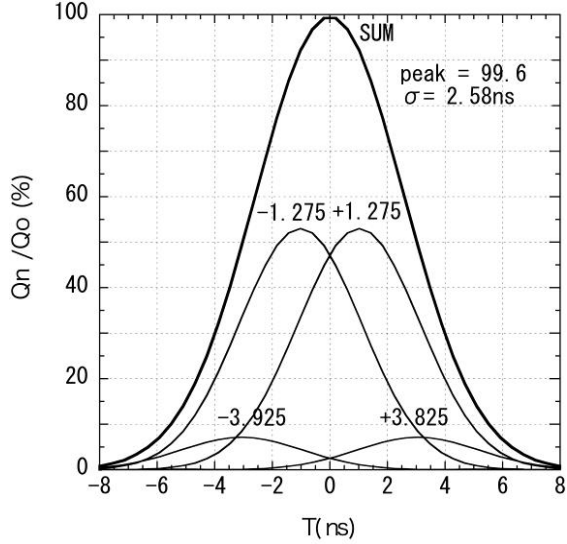

Fig. 14. Calculated output waveform for a particle incident on the gap between strips. The Gaussian labeled by SUM is the total waveform, while others are those from strips.

The very small change in the total output waveform happens because the delay-line constants and the strip pitch match well. If the delay time for the 2.55-mm pitch is 4.08 ns, which is two times larger than 2.04 ns (our delay time), the total waveform in Fig. 13 would have shoulders on both sides and that in Fig. 14 would have a flat top. If this is the case, we should have seen the pattern of the cathode strips in the position spectrum of Fig. 9.

## 7. Removal of δ ray effects by the $T_{sum}$ spectrum

The sum of the delay times $T_{sum}$ should always be constant as discussed with regard to Eq. (4). However, if the avalanche region gets wider due to the creation of δ rays, the $T_{x1}$ and/or $T_{x2}$ get(s) smaller than correct values, resulting in the decrease of the $T_{sum}$ value.

Figure 15 shows the $T_{sum}$ spectra for fission fragments with Z ~ 50 produced from the $^{238}$U + Pb reaction at 345 MeV/nucleon and light projectile fragments around $^{29}$Mg produced from the $^{48}$Ca + Be reaction at 345 MeV/nucleon, which were measured at the BigRIPS separator. The detailed setting of the BigRIPS separator for the $^{238}$U + Pb reaction was the same as the G3 setting of Ref. 7. A very small number of events are seen in Fig. 15(a) at lower channels to the left of the sharp $T_{sum}$ peak. They are due to δ rays. All particles carry δ rays when they are slowed down by passing through matter. If the energy loss of a particle due to the interaction with matter is significantly larger than that by the effect of δ rays, the latter can be ignored by setting the bias voltage of PPAC detector more or less to a low value. This is the case of the $^{238}$U + Pb reaction shown in Fig. 15(a), where fission fragments with Z ~ 50 are incident. In contrast, the $T_{sum}$ spectrum shown in Fig. 15(b) is significantly affected by δ rays, which is the case of the $^{48}$Ca + Be reaction where projectile fragments around $^{29}$Mg are incident. The $T_{sum}$ peak reveals a significant tail at lower channels. In this case the effect of the δ rays is enhanced because the energy loss of the isotopes around $^{29}$Mg is much smaller.

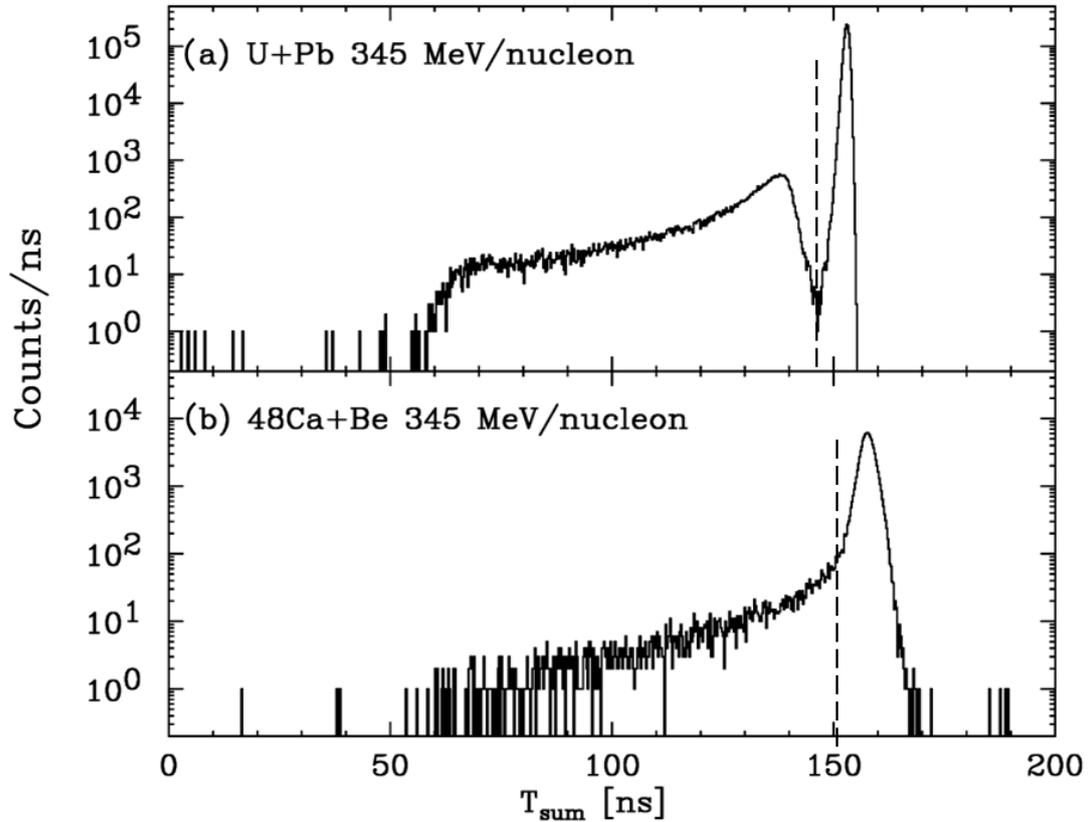

Fig. 15. $T_{sum}$ spectrum of (a) fission fragments with Z~50 produced from the $^{238}$U + Pb reaction at 345 MeV/nucleon and (b) light projectile fragments around $^{29}$Mg produced from the $^{48}$Ca + Be reaction at 345 MeV/nucleon. Dashed lines indicate the threshold to cut the events affected by δ rays. For the $^{48}$Ca + Be reaction, the threshold is set where the tail of the spectrum starts since good separation between the true events and the events affected by δ rays is not demonstrated.

Figure 16 (a) shows the two-dimensional position spectrum of a $^{29}$Mg beam produced from $^{48}$Ca fragmentation reaction on Be target at 345 MeV/nucleon. The beam should be focused as an ellipse, but the measured position spectrum has long arms pointing up, down, left, and right due to the effects of δ rays. Although the effects of δ rays are isotropic, this appears because the *X* and *Y* positions are measured in different planes of the PPAC detector. Figure 16 (b) shows the position spectrum where only the events close to the $T_{sum}$ peak are selected. In this case the correct beam profile is visible in the spectrum, because the events affected by δ rays are removed.

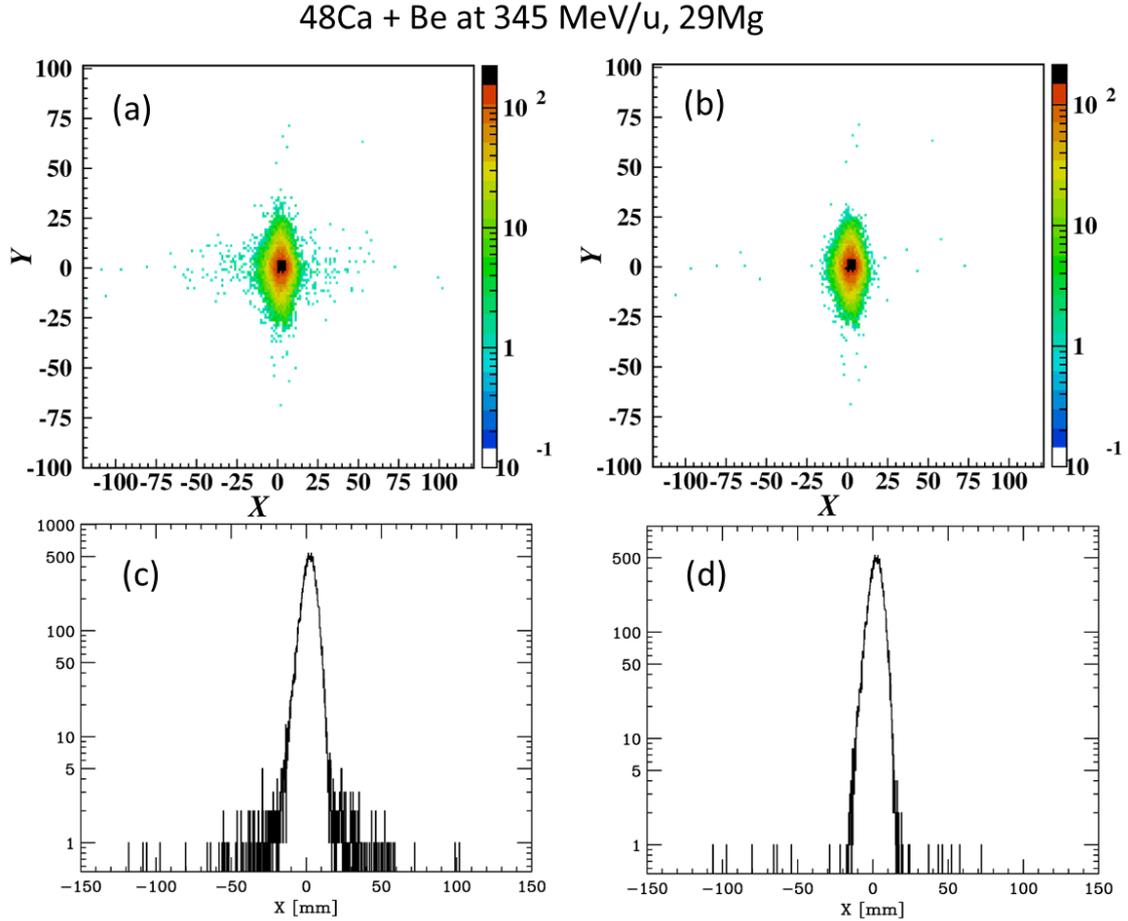

Fig. 16. Two-dimensional position spectra of a $^{29}$Mg beam produced from $^{48}$Ca fragmentation reaction on a Be target at 345 MeV/nucleon. (a) without and (b) with a $T_{sum}$ gate. The lower one-dimensional spectra (c) and (d) are those projected on the horizontal (X) plane.

## 8. Effects of using double PPAC

As schematically shown in Fig. 17, the double PPAC detector consists of two sets of the delay-line PPAC detector that are positioned in a single case, allowing twofold position measurement, and has the following advantageous features:

- One of the PPAC detectors can back up another.
- Detection efficiency can be increased by the twofold measurement. This is useful when the detection efficiency is low due to the small energy loss in the PPAC detector. For instance, if the detection efficiency of each set is 75% and one uses the events that at least one of them detects a particle, the detection efficiency increases up to 94%.
- Event loss due to the δ ray effects can be reduced.

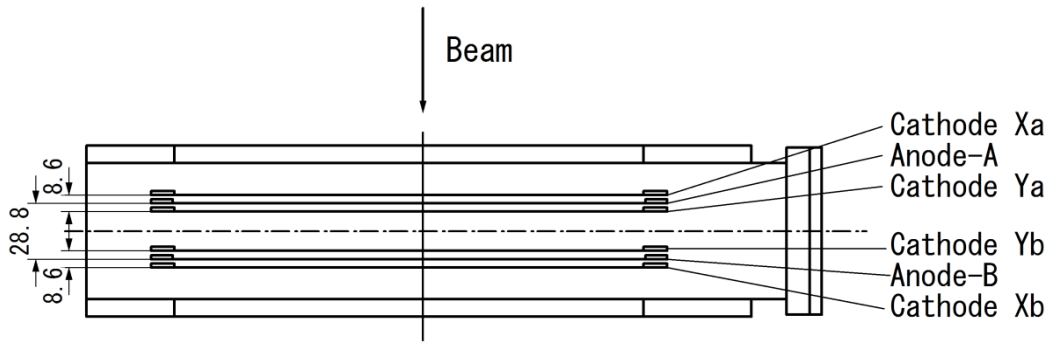

Fig. 17. A schematic drawing of the double PPAC detector. The first and second PPAC detectors with respect to the beam axis are called A-side PPAC and B-side PPAC, respectively. The dimensions are given in mm.

Figure 18 shows a two-dimensional plot of the $T_{sum}$ spectra, which was measured using fission fragments with Z around 50 produced from the $^{238}$U + Pb reaction at 345 MeV/nucleon. The horizontal axis represents the X-axis $T_{sum}$ spectrum of the A-side PPAC in a double PPAC detector, while vertical axis corresponds to the X-axis $T_{sum}$ spectrum of the B-side PPAC. As shown in Fig. 17, the A-side and B-side represent the first and second sets in a double PPAC detector, respectively. As labeled in Fig. 18, the events around the cross point of the horizontal and vertical lines are those not affected by δ rays with respect to both A- and B-sides. The events around the horizontal line indicate the case where only the A-side is affected. Those around the vertical line correspond to the case where only the B-side is affected by δ rays. Therefore, events located around the horizontal and vertical lines in the plot can be used for the position measurement. On the other hand, the events surrounded by the horizontal and vertical lines, as labeled in Fig. 18, are those affected by δ rays with respect to both A- and B-sides, corresponding to error data. The possibility that one side is affected by δ rays is 6% in this measurement. The possibility that both sides are simultaneously affected decreases significantly to 0.36%.

This difference may not matter much if the measurement is conducted at only one focus point. However, it becomes very important if several PPAC detectors are used simultaneously. For instance, four PPAC detectors become necessary if the position and angle are measured at two foci. In this case a total of eight position measurements are simultaneously made since each PPAC detector measures the *X* and *Y* positions. The total detection efficiency becomes 61% if the eight measurements have individual detection efficiencies of 94%. However, the total detection efficiency increases to 97% if the individual efficiency is 99.64%, corresponding to the case when the double PPAC detectors are employed. This improvement is very significant.

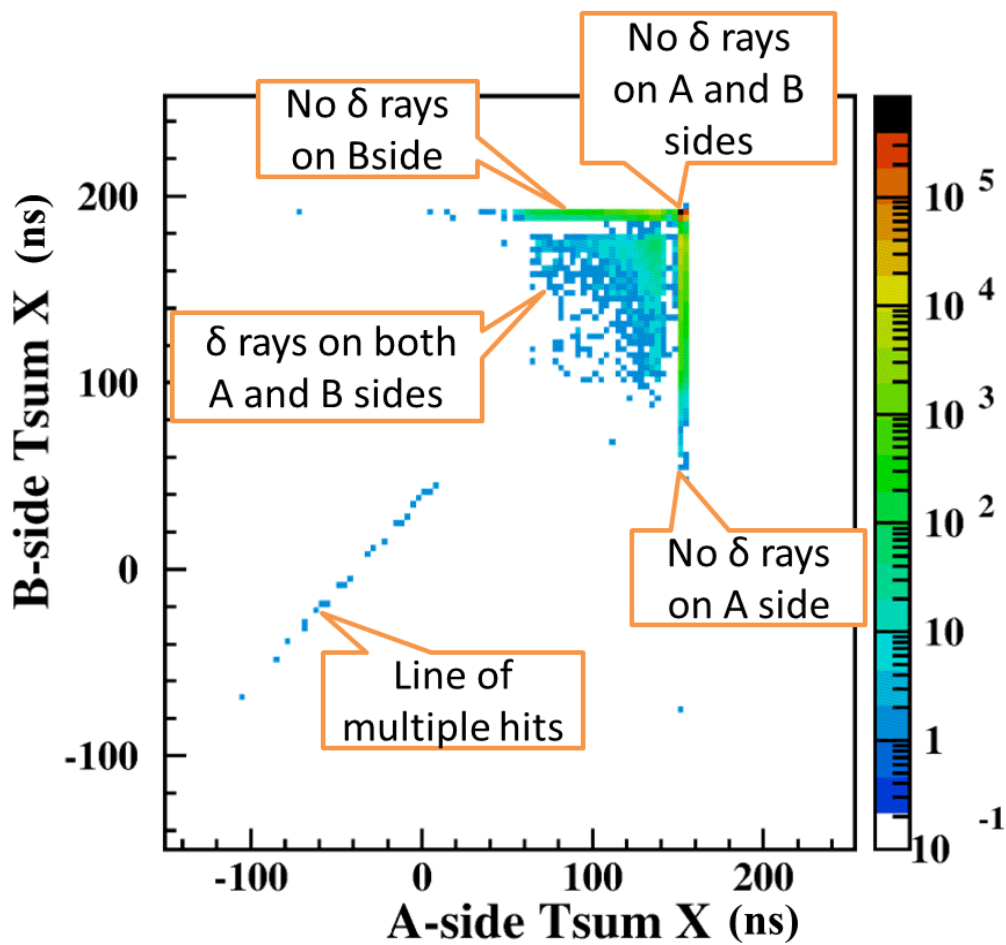

Fig. 18. Two-dimensional *X*-axis $T_{sum}$ plot for a double PPAC detector.

In the case of BigRIPS separator, two double PPAC detectors are placed at each of the four focuses in order to measure the trajectories of RI beams, while RI beams are being delivered. A total of eight PPAC detectors are used simultaneously all the time.

The events on the 45-degree line in Fig. 18 are those due to multiple hits, where two or more particles were incident during the propagation time of the delay line. Since the maximum propagation time is approximately 200 ns for 240 mm electrodes, events of multiple hits are scarce; only 2% of the events are associated with the multiple hits even if an event rate reaches 100 kcps.

## 9. Transmission of PPAC signals

The BigRIPS separator forms a 77-m long beam line. It is 126 m long if the ZeroDegree spectrometer [4] is included. Each focal point in the BigRIPS separator and ZeroDegree spectrometer is separated by about 10 m from the next. Accordingly signals from the PPAC detectors must be transmitted nearly 100 m to the data acquisition room.

The use of a coaxial cable does not allow accurate transmission, because the frequency range is narrow and sharp waveforms get distorted and attenuated. However, the use of optical fibers allows the signal transmission without distortion, because optical fiber cables have a frequency range width of 2.5 GHz [19].

Figure 19 shows the input and output waveforms of a coaxial cable and an optical fiber cable that was measured using an oscilloscope. The large, sharp waveform of the input signal is significantly broadened and the pulse height becomes small with the coaxial cable, whereas almost no change in the amplitude and rise is obtained with the optical fiber cable.

The optical fibers are suitable for use as a signal cable for the PPAC detector, because the attenuation of a few hundreds of meters long optical fibers is quite small and almost negligible (30 km, 6 dB). Furthermore the optical fiber cable does not pick up any noise.

Figures 20 and 21 show the circuit structure of the optical transmitter and receiver that we developed at RIKEN, respectively. The former uses laser diode SLT4210-DS (Sumitomo Electric Industries; infrared light, wave length= 1.31 μm). The latter uses photo diode SLT4260-CS (Sumitomo Electric Industries) with a built-in optical isolator. Despite the low cost, the time jitter of the system is very small around 0.92 ps, the noise level is as small as 1 mVp, and the dynamic range is also excellent approximately 1000 mVp.

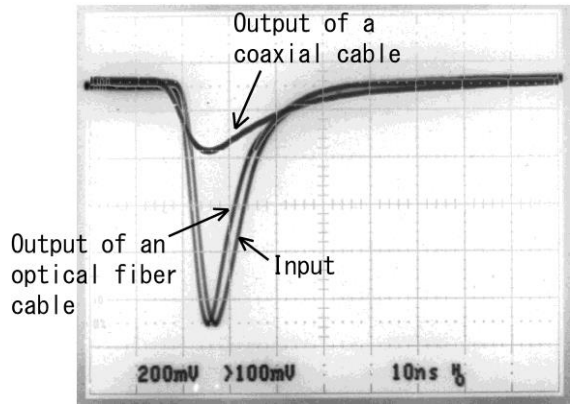

Fig. 19. Input and output waveforms of a coaxial cable and an optical fiber cable measured using an oscilloscope.

Although the transmission of timing signals (logic signals) is good enough for the position measurement using the PPAC detectors, the design allows the transmission of their analog signals. This is useful to monitor the operating conditions of the PPAC detectors.

We also use this signal transmission system for transmitting analog signals from plastic scintillation counters which is used to measure the time of flight of RI beams. The use of optical fiber cables allows insulation of the electrical ground between the experimental hall and the data acquisition room, which is advantageous to noise reduction.

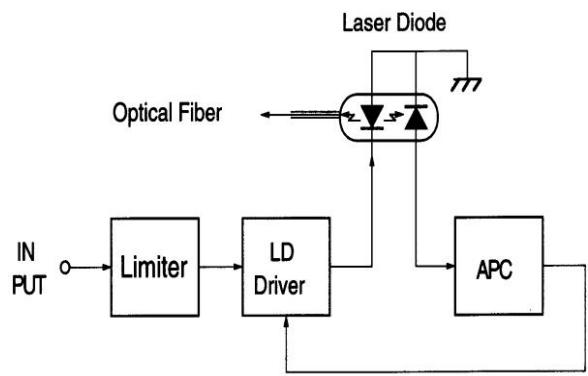

Fig. 20. Circuit structure of the optical transmitter. LD and APC stand for Laser Diode and Automatic Power Control, respectively. The chosen laser diode is SLT4210-DS, Sumitomo Electric Industries, Ltd. (infrared light, wave length: 1.31 μm)

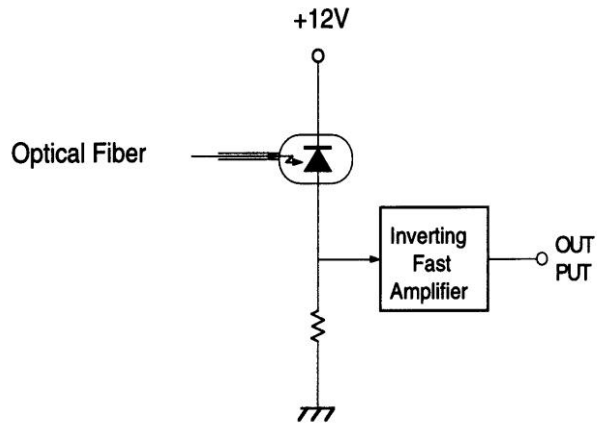

Fig. 21. Circuit structure of the optical receiver. The chosen photo diode is SLT4260-CS, Sumitomo Electric Industries, Ltd. with a built-in optical isolator.

Figure 22 shows a photograph of our optical transmitter and receiver NIM modules along with an optical fiber cable.

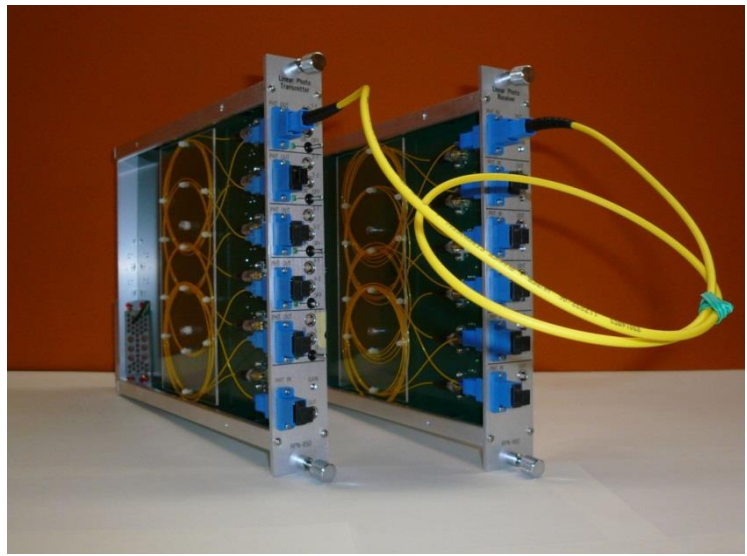

Fig. 22. Photograph of the optical transmitter and receiver NIM modules.

**10. Counter gas**

Discharges between the electrodes tend to happen more in PPAC detectors as compared with proportional counters, because the avalanche is generated immediately after the counter gas is ionized by charged particles. Therefore we cannot use proportional (PR) gas consisting of Ar 90% and $CH_4$ 10% for the PPAC detectors.

Instead, quenching gas, such as isobutylene ($C_4H_8$), isobutene ($C_4H_{10}$), isooctane ($C_8H_{18}$), hexafluoroethane ($C_2F_6$) and perfluoropropane ($C_3F_8$), are used as a counter gas of the PPAC detectors. In other gas detectors, these are mixed with main gas to prevent discharges. Among them we have investigated the characteristics of the isobutene (ion electron creation energy W = 23.7 eV), isooctane (W = 22.9 eV) and perfluoropropane (W = 34.4 eV) for our PPAC detectors.

As far as output signals are concerned, perfluoropropane gas offers the most suitable performance for our delay-line PPAC detectors. The rise time of the anode signal is measured to be as fast as 3.9 ns in this case, while those for the isobutene and isooctane gas are 4.2 and 5.8 ns, respectively. However a higher anode voltage should be applied when perfluoropropane is used: e.g. it is ~100 V higher than for isobutene. This tends to cause discharges if the gas pressure exceeds 30 torr, resulting in damage to both anodes and cathodes. Furthermore perfluoropropane is a greenhouse gas.

The use of the isobutene allows stable operation with relative low voltages at a wide range of gas pressure from ~6 to ~50 torr. Isooctane has twice the molecular mass of isobutene and a small W value, allowing production of more primary electrons. However its handling is difficult, because it is a liquid at room temperature. Perfluoropropane also can produce larger numbers of primary electrons since it has even larger molecular mass, but it tends to discharge as mentioned above.

Recently, in most cases we use isobutene for the BigRIPS separator and the delivery lines at RIBF, because overall it is the most favorable according to the above considerations.

**11. Discharge and material of electrodes**

The output signal from the PPAC detector gets larger exponentially with increasing electrical field between the gap, if the gas pressure is kept constant. If the output signal exceeds a threshold, a discharge happens, damaging to the cathode. We have observed that this threshold corresponds to the creation of ~1 x $10^8$ electrons and that the discharge creates approximately 30000 times more charge than the normal avalanche. We also have observed that the discharge current reaches a peak value in approximately 20 ns and ends in approximately 200 ns. Although this discharge phenomenon is not well understood yet, we infer that the avalanche generated along a particle trajectory grows too large, causing a streamer discharge.

The occurrence of the discharge depends on the material of electrodes. We have investigated gold, aluminum, copper, and silver for the cathode so far. In the case of gold, we found that the discharge often recurs once the electrode surface is damaged by the discharge. This can be ascribed to a protuberance created on the surface, according to our observation using a microscope. In the case of aluminum, the protuberance is not created on the surface, because aluminum evaporates when the discharge happens. However the aluminum electrode is likely to discharge if it ever happens. In the case of copper, similarly to aluminum, the discharge evaporates copper and the protuberance is not created on the surface. However the copper electrode does not tend to discharge even if it is damaged by the discharge. In these respects the copper electrode is more stable than the gold and aluminum ones. However oxidization of copper is a drawback, because it reduces the conductivity of the electrode. Silver electrodes are not oxidized in air. We recently tested a silver electrode and have found that it is more tolerant of the discharge than the aluminum and copper electrodes.

We currently use the aluminum or copper electrodes at RIBF.

## 12. Performance measurement using RI beams produced by BigRIPS separator

Since March 2007 we have used the double PPAC detectors to tune the production and delivery of RI beam. Two double PPAC detectors are placed at each focus in the BigRIPS separator and subsequent RI-beam delivery lines. These PPAC detectors are used to measure the profile and phase space of RI beams, allowing the diagnostics of RI beams and the ion-optical tuning of the separator and delivery lines. The trajectories of RI beams are also measured at the foci in the BigRIPS separator in order to reconstruct trajectories, determining Bρ with high resolution. This high resolution Bρ determination is crucial to achieve excellent resolving power in particle identification of RI beams [6,7].

Figure 23 shows detection efficiency of the PPAC detectors for fission fragments produced in the reaction $^{238}$U + $^{9}$Be at 345 MeV/nucleon along with their *Z* versus *A/Q* particle identification plot. Here the experimental conditions and the BigRIPS separator settings correspond to those of the G2 setting described in Ref. [7]. The detection efficiency is shown as a function of the *Z* numbers of fission fragments for the PPAC detectors placed at the F7 focus in the BigRIPS separator (see the layout shown in Ref. [4]). The applied anode voltage was 1055 V and the used counter gas was $C_3F_8$ at 10 torr in this case. The PPAC detectors were tuned for fission fragments with *Z* ~40 to 50. The closed circles in the right figure of Fig. 24, denoted by 'Double', indicate the detection efficiency in the double PPAC mode in which the beam trajectory is obtained by the twofold measurement using the two double PPAC detectors. Note that all the layers in the PPAC detectors do not necessarily hit in this mode. On the other hand, the open circles correspond to the detection efficiency in the single PPAC mode in which the trajectory is obtained using only one layer in each double PPAC detector. The improvement of detection efficiency can be clearly seen in the figure, demonstrating the great advantage of using the double PPAC detectors.

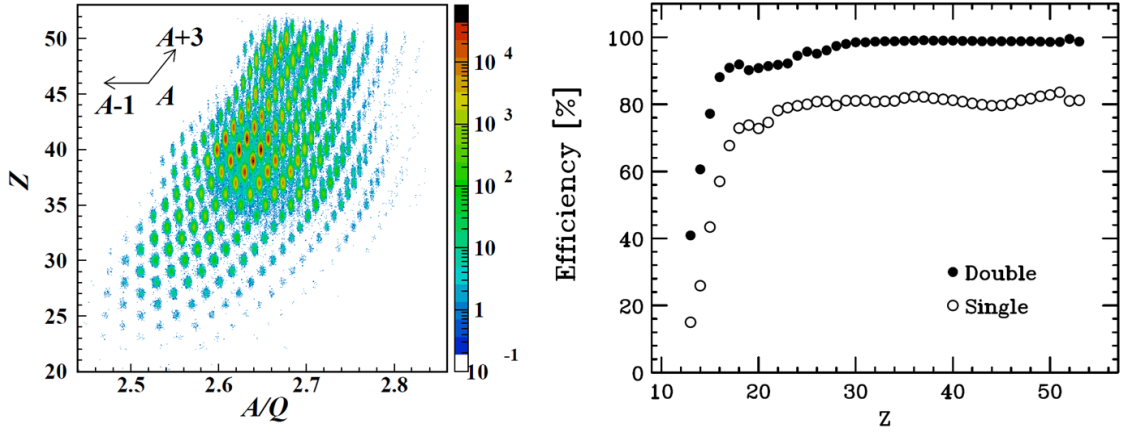

Fig. 23. (Left) $Z$ versus $A/Q$ particle identification plot for fission fragments that are produced in the reaction $^{238}$U + $^{9}$Be at 345 MeV/nucleon. (Right) Detection efficiency of PPAC detectors for these fission fragments as a function of $Z$ number. The closed and open circles indicate the detection efficiency in the double and single PPAC modes, respectively. See text.

Figure 24 shows the position resolution of a PPAC detector at the F7 focus for the fission fragments measured in the same run as in Fig. 23. As illustrated in the right of Fig. 24, in order to investigate the position resolution, we have deduced the position dispersion in the layer of interest with respect to the trajectory, which is obtained by fitting the measured positions for the rest of layers. The achieved r.m.s. resolution is 0.39 mm for fragments with $Z$ ~50.

We have obtained a similar r.m.s. position resolution for light projectile fragments which were produced in the reaction $^{48}$Ca + $^{9}$Be at 345 MeV/nucleon. The applied anode voltage was 850 V and the counter gas was $C_4H_{10}$ at 11 torr in this case. The achieved r.m.s. resolution was 0.41 mm for fragments with $Z = 12$.

In these measurements, the incident angle dependence of the position resolution was not observed since the angular distribution of the fragments was limited by the acceptance of the BigRIPS separator which was as narrow as 3 degree

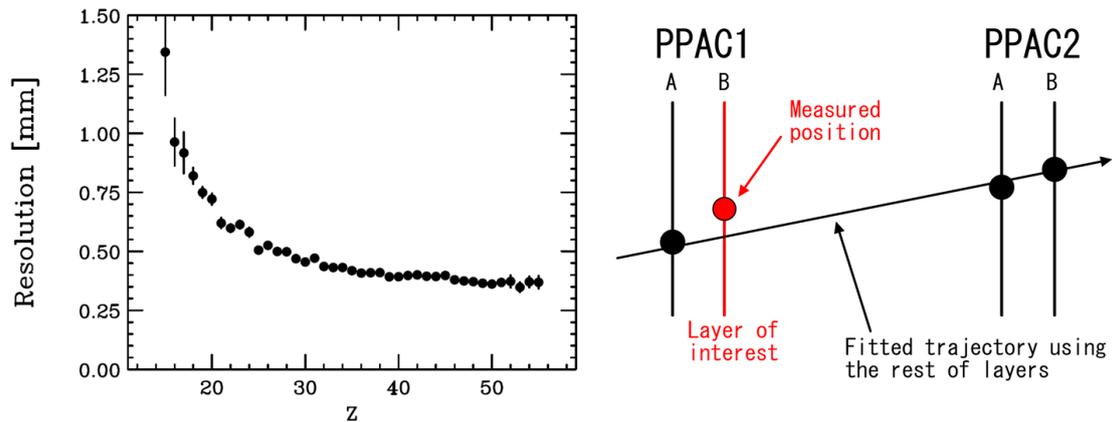

Fig. 24. (Left) Position resolution (r.m.s) of the PPAC detector as a function of Z number for fission fragments that are produced in the reaction $^{238}$U + $^{9}$Be at 345 MeV/nucleon. The experimental conditions and the BigRIPS separator settings are the same as in Fig. 23. (Right) Illustration that schematically shows how to deduce the position resolution. The position resolution is deduced for the layer of interest with respect to the trajectory that was obtained by fitting the measured positions using the rest of layers.

## 13. Summary and conclusion

In the present paper we outlined the development of the PPAC detectors at RIKEN Nishina Center RI Beam Factory, including their operating principles, design and specifications, characteristics and performance, issues, and signal transmission. They are what we call double PPAC detector, by which twofold position measurement is made possible. The PPAC detectors adopt a delay-line readout scheme which is based on our independently developed (proprietary) technology. All the foci in the BigRIPS fragment separator and the subsequent RI-beam delivery line such as the ZeroDegree spectrometer are equipped with two PPAC detectors, allowing trajectory measurement with high detection efficiency and reasonably good position resolution. They have been used for diagnostics of RI beams and the ion-optical tuning. The trajectory measurement using PPAC detectors has been crucial in high-resolution Bρ determination and hence in achieving excellent particle identification power of the BigRIPS separator.

**Acknowledgements**
The authors would like to thank the accelerator crew at RIBF for providing us with primary heavy ion beams. They also would like to thank Dr. Y. Yano, RIKEN Nishina Center, for his support and encouragement. T.K. is grateful to Dr. J. Stasko for his careful reading of the manuscript.

Corresponding authors:
Hidekazu Kumagai (kumagai@ribf.riken.jp) and Toshiyuki Kubo (kubo@ribf.riken.jp)